%% file: usenix2019_v3.1.tex
\documentclass[letterpaper,twocolumn,10pt]{article}
\usepackage{usenix2019_v3}

\usepackage{tikz}
\usepackage{amsmath}
\usepackage{amsfonts}
\usepackage{wrapfig,booktabs}
\usepackage[linesnumbered,ruled,vlined]{algorithm2e}
\usepackage[font=footnotesize,caption=false]{subfig}
\usepackage{graphicx}
\usepackage{color}
\usepackage{multirow}
\usepackage{verbatim}
\usepackage{diagbox}

\usepackage{siunitx}
\usepackage{caption}
\usepackage{makecell}

\usepackage{balance}

\newcommand{\aaf}{\vspace*{-6pt}}

\usepackage{filecontents}

\begin{document}

\date{}

\title{\Large \bf Exploiting the Inherent Limitation of $L_0$ Adversarial Examples}


\author{
{\rm Fei Zuo}\\
University of South Carolina\\
fzuo@email.sc.edu
\and
{\rm Bokai Yang}\\
University of South Carolina\\
bokai@email.sc.edu
\and
{\rm Xiaopeng Li}\\
University of South Carolina\\
xl4@email.sc.edu
\and
{\rm Lannan Luo}\\
University of South Carolina\\
lluo@cse.sc.edu
\and
{\rm Qiang Zeng}\\
University of South Carolina\\
zeng1@cse.sc.edu
} 

\maketitle

\input{abstract.tex}

\input{intro.tex}

\input{bckground.tex}

\input{sys.tex}

\input{eval.tex}

\input{adaptive.tex}

\input{related.tex}

\input{conclude.tex}

\section*{Acknowledgments}

We would like to thank our shepherd, Dr. Zachary Weinberg, and the anonymous reviewers for their constructive suggestions and comments. This project was supported by NSF CNS-1815144, NSF CNS-1856380, and NSF CNS-1850278.



\balance
\bibliographystyle{plain}
\bibliography{ref.bib}

\end{document}

%% file: abstract.tex
\begin{abstract}
Despite the great achievements made by neural networks on 
tasks such as image classification, they are brittle 
and vulnerable to adversarial example (AE) attacks,
which are crafted by adding human-imperceptible
perturbations to inputs in order that a neural-network-based 
classifier incorrectly labels them.
In particular, $L_0$ AEs are 
a category of widely discussed threats where adversaries are
restricted in the number of pixels that they can corrupt. 
However, our observation is that, while 
$L_0$ attacks modify as few pixels as possible, 
they tend to cause large-amplitude perturbations to 
the modified pixels. We consider this as an inherent 
limitation of $L_0$ AEs, and thwart such attacks  
by both detecting and rectifying them. 
The main novelty of the proposed detector is that 
we convert the \emph{AE detection problem} into a \emph{comparison problem} 
by exploiting the inherent limitation of $L_0$ attacks. 
More concretely, given an image $I$, 
it is pre-processed to obtain another image $I^\prime$. 
A Siamese network, which is known to be effective in comparison, 
takes $I$ and $I^\prime$ as the input pair to determine whether $I$ is an AE. 
A trained Siamese network 
automatically and precisely captures the discrepancies 
between $I$ and $I^\prime$ to detect $L_0$ perturbations. 
In addition, we show that the pre-processing technique, \emph{inpainting}, used for detection 
can also work as an effective defense, 
which has a high probability of removing the adversarial influence of $L_0$ perturbations. Thus, 
our system, called  \textsc{AEPecker}, demonstrates not only high AE detection accuracies, but also
a notable capability to correct the classification results.

\end{abstract}

%% file: intro.tex
\section{Introduction}
Recent years have witnessed tremendous success of neural networks in a variety of fields, 
such as object detection~\cite{ren2015faster}, motion tracking~\cite{wang2013learning}, face recognition~\cite{parkhi2015deep,wen2016discriminative}, and code analysis~\cite{zuo2019neural,redmond2018cross,luo2016solminer}.
Despite these great achievements, they are vulnerable to adversarial examples (AEs).
Szegedy et al.~\cite{szegedy2014intriguing} analyze the robustness of neural networks when facing adversarial attacks, and show that deep learning systems are sensitive to small adversarial perturbations. A neural-network-based classifier thus can be misled by AEs and generate incorrect
classification results. Many image AE generation methods have been proposed and multiple 
off-the-shelf tools are available~\cite{goodfellow2014explaining, papernot2016limitations, kurakin2016adversarial, carlini2017towards}. 

The adversarial perturbations in an image AE are usually subtle
in order to be human-imperceptible.
To quantitatively describe such perturbations, 
$L_p$ norms are usually used to measure the discrepancy between an original 
benign image $I_o$ and its corresponding AE $I_a$. 
According to the value of $p$, the mainstream AE generation algorithms 
can be categorized into three families, i.e., $L_0$, $L_2$ and $L_{\infty}$ attacks. 
Informally, $L_0$ measures the number of modified pixels, 
$L_2$ the Euclidean distance 
between the two images, and $L_{\infty}$ the 
largest modification among the pixels. 
Note that our work focuses on $L_0$ AEs, a category of attacks widely considered by previous works~\cite{papernot2016limitations, carlini2017towards, xu2017feature,ma2019nic}.


To defeat attacks based on AEs, both detection and defensive techniques attract the research community's attention.
Given an input image, the \emph{\textbf{detection}} system outputs whether it is an AE, so that the target neural network can reject those adversarial inputs.
A \emph{\textbf{defense}} technique,
given an AE, helps the target neural network make correct 
prediction by either rectifying the AE or fortifying the classifier itself.

Many AE detection methods~\cite{monteiro2018generalizable, liang2018detecting, bagnall2017training} and 
defense techniques~\cite{das2018shield,xie2017mitigating,liao2018defense} have been proposed.
However, prior methods either are not very effective in handling $L_0$ AEs or omit discussing them. 
For example, feature squeezing~\cite{xu2017feature} is capable of detecting $L_0$ AEs. 
However, He et al.~\cite{he2017adversarial} have shown that feature squeezers, 
either single or joint, are not resilient to adaptive attacks.
Previous work even argues that 
it is challenging to recover the correct classification of 
$L_0$ AEs by input transformation, as ``\emph{it is very difficult to properly reduce the 
effect of the heavy perturbation}''~\cite{liang2018detecting}.

We identify two characteristics of $L_0$ AEs.  
By exploiting the two characteristics, we build a detector based on a very simple architecture that
achieves a high detection accuracy. Moreover, 
a pre-processor based on these observations can effectively rectify $L_0$ AEs to
recover the correct classifications. 

The first characteristic is that it limits the number of modified 
pixels, but not the amplitude of pixels. 
Thus, $L_0$ attacks  tend to introduce large-amplitude perturbations, 
especially for targeted attacks that aim to achieve an attacker-desired output from a neural network. 
Second, as $L_0$ attacks try to modify as few pixels as possible, 
the optimization-based AE generation process tends to result in 
altered pixels that scatter in the image.
In other words, those corrupted parts are mostly small and isolated regions. 
Both characteristics are verified by our experiments.

We accordingly propose
a novel AE detection method. 
The main novelty is that we convert the \emph{AE detection
problem} into a \emph{comparison problem}. Specifically,
the architecture of the detector uses 
a Siamese network~\cite{bromley1994signature}, which is known to
be powerful in comparison. Given an image $I$,
it is processed by a pre-processor to obtain another image $I^\prime$.
The Siamese network takes $I$ and 
$I^\prime$ as the inputs and outputs whether $I$ is
an AE. The advantage of the design is that the Siamese
network is able to automatically and precisely 
capture the discrepancies
between the two inputs for AE detection. 

Another advantage is that
the pre-processor used for AE detection can also work 
as an effective defense by removing the 
influence of the adversarial perturbations. 
Specifically, we propose an \emph{inpainting}-based algorithm to process images, 
where \emph{inpainting} refers to the process of \emph{reconstructing} the lost or corrupted 
parts of an image. 
The inpainting techniques are a fruitful sub-field in the area of digital image processings~\cite{shen2002mathematical,telea2004image,mairal2007sparse}, which have been widely used in practice.
As we will show in Section~\ref{sec:eval_preproc}, inpainting is more
effective at eliminating the heavy perturbations created by $L_0$ attacks
than previous defenses.



\begin{figure}[!t]
\vspace{6pt}
\hspace{-3pt}
\centerline{\includegraphics[scale=0.43, trim=0 0 0 0,clip]{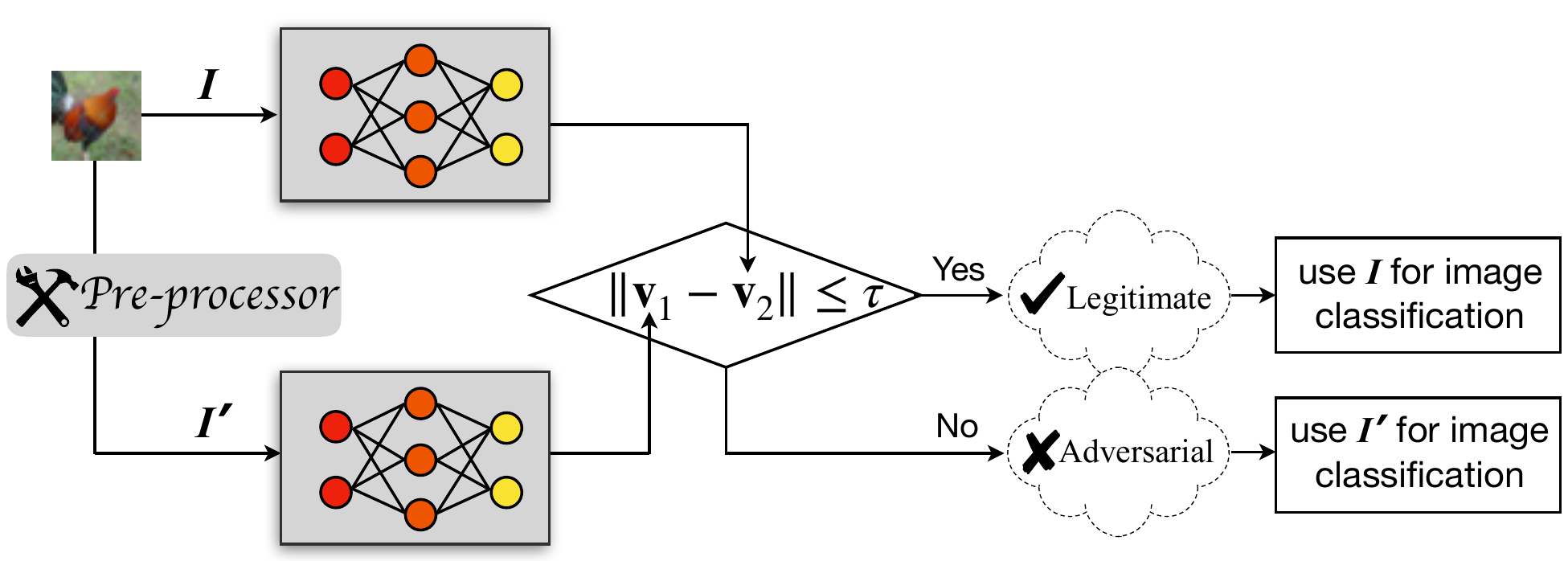}}
\caption{An architecture of \textsc{AEPecker}. If $I$ is detected as an $L_0$ AE,  $I'$ is used for image classification as a defense.}\label{fig:sys_design}
\vspace{-6pt}
\end{figure}

We implement a system \textsc{AEPecker} 
to demonstrate the advantages aforementioned and the
weakness of $L_0$ attacks. 
The system architecture is shown in Figure~\ref{fig:sys_design}.
After inputting an image $I$ to a pre-processor $\mathcal{P}$, we obtain another image $I^\prime$. Then, the Siamese network predicts whether $I$
is adversarial  by taking $\langle I,I^\prime \rangle$ as the input pair. 
If $I$ is detected as an $L_0$ AE, then we regard $I^\prime$ 
as a rectified image and use it to replace $I$ in subsequent image classification
for the defense purpose. 

We have evaluated 
our system in terms of its detection and defense capabilities using the popular image datasets CIFAR-10 and MNIST. Two leading $L_0$ AE generation methods, JSMA~\cite{papernot2016limitations} and 
CW-$L_0$~\cite{carlini2017towards}, are both considered in the evaluation.
In the case of CIFAR-10 (we have similar results for MNIST), 
the evaluation results show that (1) 
the detection rate on the CW-$L_0$ and JSMA attack  
is 97.1\% and 99.7\% respectively, both with a low false positive rate; (2) the proposed system has outstanding \emph{transferability}, as a detector trained only with 
JSMA AEs can detect CW-$L_0$ AEs with a high detection 
rate (99.4\%),
and vice versa; (3) the detection is also \emph{attack-target-model agnostic} (model agnostic, for short), since in the aforementioned experiments CW-$L_0$ AEs
and JSMA AEs actually target different image classification models; and (4) our defense method recovers the 
classification accuracy from 0\% (when classifying those successful AEs) to 87.3\% for CW-$L_0$, and from 0\% to 96.1\% for JSMA,
and meanwhile, has a very small impact on benign images. 

Moreover, in order to illustrate the effectiveness of the Siamese network in detecting AEs,
we experiment to use a preprocessing technique, \emph{bit depth reduction}, that is known
to be weak. Feature squeezing~\cite{xu2017feature} used it as one of the pre-processors 
and obtained an AE detection rate 4.1\%. In contrast, the Siamese network plus the weak preprocessing technique achieves 99.6\%, which demonstrates the unique advantage of the Siamese architecture in
detecting AEs.

The key contributions of our work include: 
\begin{itemize}
  \item We point out the inherent characteristics of
$L_0$ AEs, which typically contain high-amplitude
perturbations to very few and isolated pixels, and propose to
exploit them to develop detection and
defense techniques. 
  
  \item We convert the \emph{$L_0$ AE detection
problem} into an \emph{image comparison problem},
and propose to use a Siamese network to
automatically extract the subtle discrepancies of the
input pair as features for the AE detection.
The detector demonstrates multiple prominent 
  strengths, such as transferability across attacks
  and being attack-target-model agnostic (so the detector keeps
  effective across attack methods and target classifiers). 

  \item We propose an effective \emph{inpainting}-based 
  defense against $L_0$ perturbations, which
  can recover the correct classification at a 
  high probability. To the best of our knowledge,
  this defense method achieves the highest accuracy when dealing
  with $L_0$ AEs.
  
  \item Adaptive attacks that try to bypass our detection 
  are considered and evaluated. The evaluation results show
  that our system is resilient to them. 
  
\end{itemize}

The rest of the paper is organized as follows. 
First, we briefly introduce some background about $L_0$ AE generation methods in Section~\ref{sec:ae_generation}. 
Section~\ref{sec:sys} describes our system architecture and design. 
Then, we present the evaluation design and results in Section~\ref{sec:eval}. 
We also empirically evaluate the resilience of the proposed technique under
adaptive attacks in Section~\ref{sec:adapt}.
The related work is then reviewed in Section~\ref{sec:relate_work}. 
We finally discuss the limitations of our work and draw conclusions in Section~\ref{sec:limit} and~\ref{sec:conclude}, respectively.

%% file: bckground.tex
\section{Adversarial Example Generation}\label{sec:ae_generation}


Adversarial examples are carefully crafted inputs intended to fool 
artificial intelligence systems to output incorrect labels. 
The term \emph{adversarial example} can be formally defined as following. For a pre-trained neural network $f$, let $x$ be an original image. An adversarial example $x^{adv}$ is such an intentionally designed input by attackers which can guide the model $f$ to make an incorrect prediction. Moreover, to hide the adversarial perturbation, the generation of $x^{adv}$ is equivalent to solve the following constrained optimization problem:
\begin{align*}
 \min\limits_{x^{adv}}\ &\Vert x^{adv} - x\Vert_p \\
\mathrm{s.t.}\ \ \bar{y} &= f(x^{adv}) \\
y &= f(x) \\
y &\neq \bar{y}
\end{align*}
where $y$ and $\bar{y}$ are respectively the prediction results of feeding $x$ and $x^{adv}$ to $f$, 
and $\| \cdot \|_p$ denotes the $L_p$-norm.

Based on the value of $p$, three
metrics exit, i.e., $L_0$, $L_2$, and $L_{\infty}$, which are usually used to measure human’s perception of visual difference Specifically, $L_0$ measures how many pixels are modified at the corresponding positions in the resulting image; 
$L_2$ represents the Euclidean distance between the two images; and $L_{\infty}$ measures the maximum difference for all pixels at the corresponding positions in the two images.

Depending on the manner of how $\bar{y}$ misleads a pre-trained classifier, adversarial attacks to neural networks can be categorized as either targeted or non-targeted. The aim of non-targeted attacks is to make the image be classified as any arbitrary class except the true one. By contrast, in targeted attacks the prediction result will be misguided to a specific class different from the correct one and desired by the attacker. 

In this study, we focus on the discussion of $L_0$ AE attacks, where JSMA and CW-$L_0$ are two widely used and representative $L_0$ AE generation methods. Next, we will describe these AE generation methods briefly.

\subsection{Jacobian Saliency Map Attack (JSMA)}

The JSMA is a targeted attack based on a greedy iterative idea proposed by Papernot 
et al.~\cite{papernot2016limitations}. It takes $L_0$ distance minimization as the optimization target, that is, the number of pixels that can be updated in the original image is bounded. 
To determine which pixels will be manipulated, the authors introduce the concept of \textit{saliency map} which provides an adversarial saliency score for each pixel.
One single pixel that possesses a higher adversarial saliency score usually has more impact on misleading the target model to predict a specific label desired by attackers.
Thus, the attacker only manipulates those pixels that have high 
adversarial saliency scores in each iterative step based on a greedy strategy. 
The adversarial saliency score for
each pixel is calculated as:

\aaf
$$
    x^{adv}_{i,t} = x_{i,t}+
    \begin{cases}
    0,\quad\quad \mathrm{if}\ \frac{\partial f_t(x)}{\partial x_i}<0\ \mathrm{or}\ \sum\limits_{j\neq t}\frac{\partial f_j(x)}{\partial x_i}>0\\
    \frac{\partial f_t(x)}{\partial x_i}|\sum\limits_{j\neq t}\frac{\partial f_j(x)}{\partial x_i}|,\quad\quad\quad\quad \mathrm{otherwise}
    \end{cases}
$$
where $i$ denotes the $i$th pixel in the image, and $f_j$ is the prediction value 
of the neuron $j$ in the target model's output layer.




\subsection{Carlini \& Wagner Attack (CW)}

CWs are a group of targeted AE generation methods developed by Carlini and 
Wagner~\cite{carlini2017towards}. 
There are three types of CW attacks that use distance metrics $L_0$-, $L_2$- and $L_{\infty}$-norm, respectively. 
In this work, we 
focus on the first type of CW attacks, where $L_0$-norm is used as the distance metric during the construction of AEs. In the following presentation, We refer to such a CW attack as CW-$L_0$.

Some notable features are developed by the authors which make CW attacks very effective. First, to compute the loss in gradient descent, the algorithm does not directly use final prediction given by the target model; instead, a logit function is used which plays a key role in the resilience improvement of the attack against 
defensive distillation~\cite{papernot2015distillation}. 
Second, this algorithm maps the target variable to a space of the inverse trigonometric function; as a result, the optimization problem is suitable to be computed by a modern solver such as Adam~\cite{kingma2014adam}. Finally, a particular constant is designed to adjust the relative importance 
between perturbations and the misclassification; through this, a fine-grained trade-off is enabled.
These techniques ensure that the CW method can generate superior adversarial examples with minimized perturbations.

%% file: sys.tex
\section{System Design}\label{sec:sys}

The proposed system consists
of a Siamese network (Section~\ref{sec:detector}) 
which determines whether an input image is an $L_0$ AE, 
and a pre-processor (Section~\ref{sec:defense-component}) which also can be used as a defense component
to correct the classification under the
existence of $L_0$ AEs. 
Note that the pre-processor has a very small impact on benign images; thus it can be used as a defense component independently
without relying on detection.

\subsection{Pre-processor}
\label{sec:defense-component}

The pre-processor adopted in our system is designed to reduce adversarial noises while preserving the features in images to reduce false positives. From this perspective, the proposed pre-processor can also be deployed as a defense against $L_0$ attacks.

Intuitively, failing to limit the amplitude of those altered pixels in the 
images will result in outlier pixels. Previous work ~\cite{liang2018detecting} emphasizes 
that it  is challenging to get rid of the effect of those heavy perturbations. 
However, we argue the outlier pixels can be fixed by 
applying a processor based on \emph{inpainting}. 
In image processing, the term ``inpainting'' refers to the process of reconstructing lost 
or corrupted regions of image data (or to remove small defects). 
Our idea is to treat those 
outliers as small corrupted regions, and the inpainting technique exactly 
meets the need for eliminating the $L_0$ noise.

\begin{figure}[!t]
\hspace{-5pt}
\centerline{\includegraphics[scale=0.24]{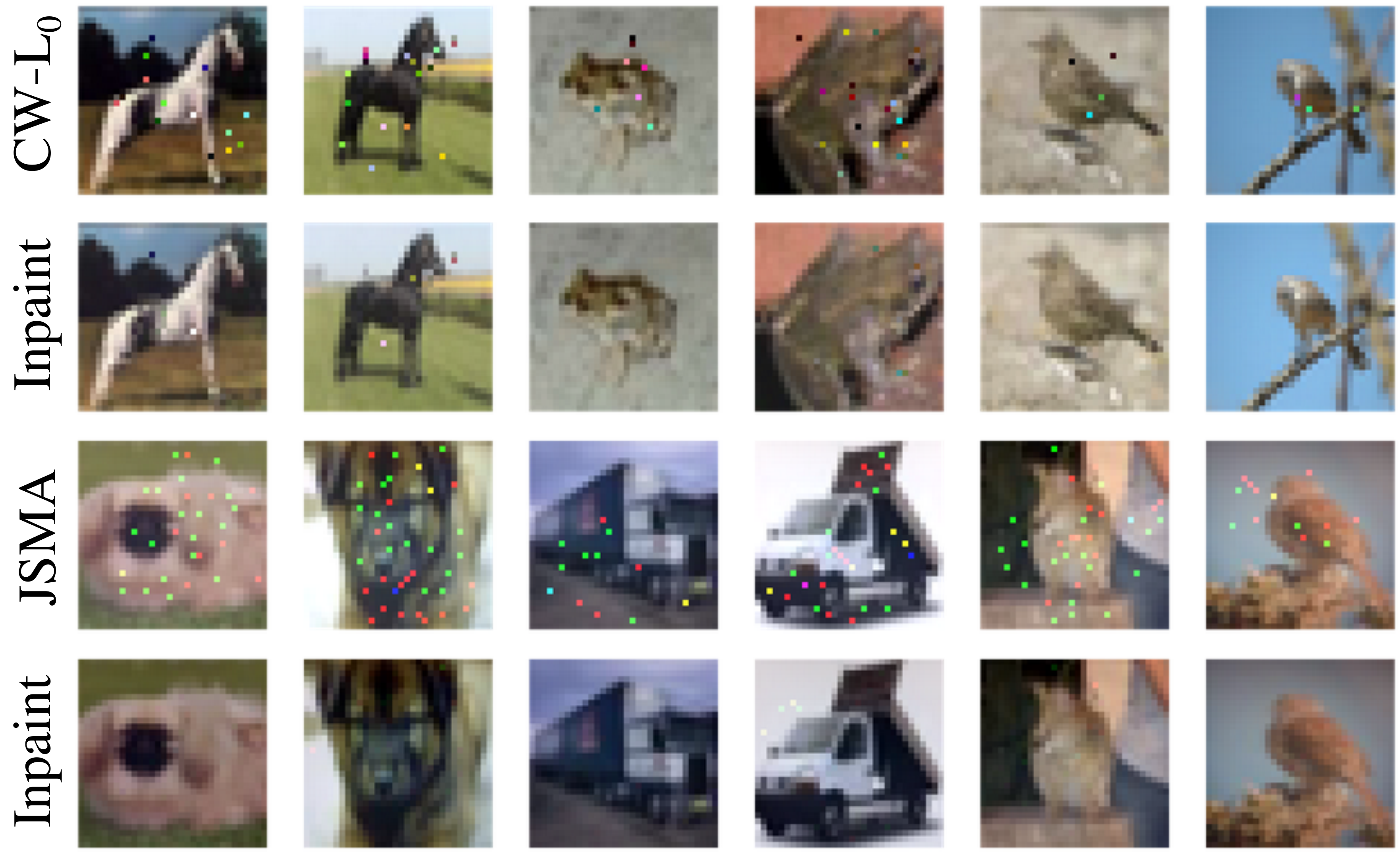}}
\caption{Defense based on inpainting. 
The first and third rows show the CW-$L_0$ and JSMA attack applied to CIFAR-10 images, respectively. The second and fourth rows show the corresponding resulting images after inpainting.}\label{fig:inpaint}
\end{figure}

In detail, we observe that those $L_0$ perturbations manifest themselves 
visually as salient noises. 
A mask to determine which pixels should be reconstructed can help identify these cases. 
When inspecting the pixel intensity in different color channels
(e.g., the R, G, B channels for color images), 
 for an altered pixel, it is highly possible that one \textit{extreme value} can be observed in at least one channel.
For example, an original pixel 
is represented as an intensity vector [0.32, 0.56, 0.62], where all the values are normalized. After corrupting by 
the $L_0$ attack, it becomes [0.33, 0.55, 0.96], whose B channel has an \textit{extreme value} 0.96. We define a value as \textit{extreme} if it is either smaller than an upper bound $\alpha$ or larger than an lower bound $\beta$. Thus, to obtain such a mask, we first locate all pixels of which the intensity are exceptional at least one channel. 
Meanwhile, we noticed that such pixels that achieve \textit{extreme values} in all 
of the three channels are often the bright parts such as the sky in a natural image.
Therefore, we use a parameter $\gamma$ to help filter out such pixels in color images. 
According to our observation, we choose $\gamma=0.7$ as an empirical value.
Lines 4-10 in Algorithm~\ref{alg:inpaint} show the procedures to initially create the mask.

\input{alg.tex}

In addition, considering that the number of altered pixels only occupies a small portion of the image, the possibility that most of the altered pixels will assemble to form a connected region is very low. Consequently, to further exclude those unlikely candidates, we will remove those relatively large connected regions from the mask. Specifically, we use a \textit{structuring element} $\mathcal{E}$ to describe a connected region with the specified size and shape. If a connected region is larger than $\mathcal{E}$, we will exclude such region from the mask, as Lines 11-13 in Algorithm~\ref{alg:inpaint} show, where $\mathit{N}(\cdot)$ denotes a connected neighborhood.

We thus independently 
produce an inpainting mask 
for each channel of a color image. We then take advantage of the inpainting method 
proposed in~\cite{telea2004image} to restore those deteriorated pixels for each channel, 
as Lines 14-15 in Algorithm~\ref{alg:inpaint} show. 
Figure~\ref{fig:inpaint} displays some concrete examples applying 
Algorithm~\ref{alg:inpaint} with $\alpha=0.2$, $\beta=0.8$ on CIFAR-10. The resultant images in the even numbered row show that the adversarial perturbations are almost completely eliminated. We will provide more detailed experimental results to demonstrate how our defense influences 
the effectiveness of $L_0$ attacks in Section~\ref{sec:eval}.

The algorithm for gray images is very similar to Algorithm~\ref{alg:inpaint}, but we only need to consider one channel rather than three. Thus, we can consider the algorithm for gray images as a special case of the algorithm for color images.

\vspace{3pt}
\noindent \textbf{\emph{Parameters selection.}}
At beginning, our algorithm normalizes the value of all input pixels, such that their values are in the range of $[0, 1]$. (1) $\alpha$ is the upper bound of extremely small values; thus, the value of $\alpha$ should be
small (e.g. less than 0.2). (2) $\beta$ is the lower bound of extremely large values; thus, it should be relatively large (e.g. 0.7 at least).
Different parameters settings slightly affect the effectiveness of rectifying AEs. We show the experiment results in Section~\ref{sec:eval_preproc}. 
(3) In addition, as aforementioned, we use a parameter, $\gamma$,  to help filter out the normal bright parts in a natural image. The term \textit{atmospheric light} refers to those pixels, which has been discussed in detailed in the field of image processing~\cite{kim2013optimized}.
Based on our experience, in our experiment (Section~\ref{sec:eval}), the value of $\gamma$ is set to 0.7.
(4) Finally, the structuring element $\mathcal{E}$ is closely related to the restoration capability of the inpainting algorithm and the size of input images. The corrupted region that can be restored by the widely used inpainting algorithms is not only a single pixel but also a small patch~\cite{shen2002mathematical,telea2004image,mairal2007sparse}.
However, as the size of patch increases, the restoration effect usually degrades. 
A recommendation size of $\mathcal{E}$ given by~\cite{telea2004image} ranges from three to ten pixels.
Note that the performance of the pre-processor has little impact on the detection accuracy of \textsc{AEPecker}, as demonstrated in our evaluation (see Section~\ref{sec:var_pro}).


\subsection{Siamese Network-Based Detector}
\label{sec:detector}

As a classic category of neural network architecture, Siamese networks~\cite{bromley1994signature} are widely applied among those tasks that involve detecting similarities or other relationships between two or more comparable thingss~\cite{zuo2019neural}. In general, a Siamese network consists of two sub-networks which share one identical architecture with the same weights. 

\begin{figure}[!t]
\centering
\centerline{\includegraphics[scale=0.40, trim=0 0 0 0,clip]{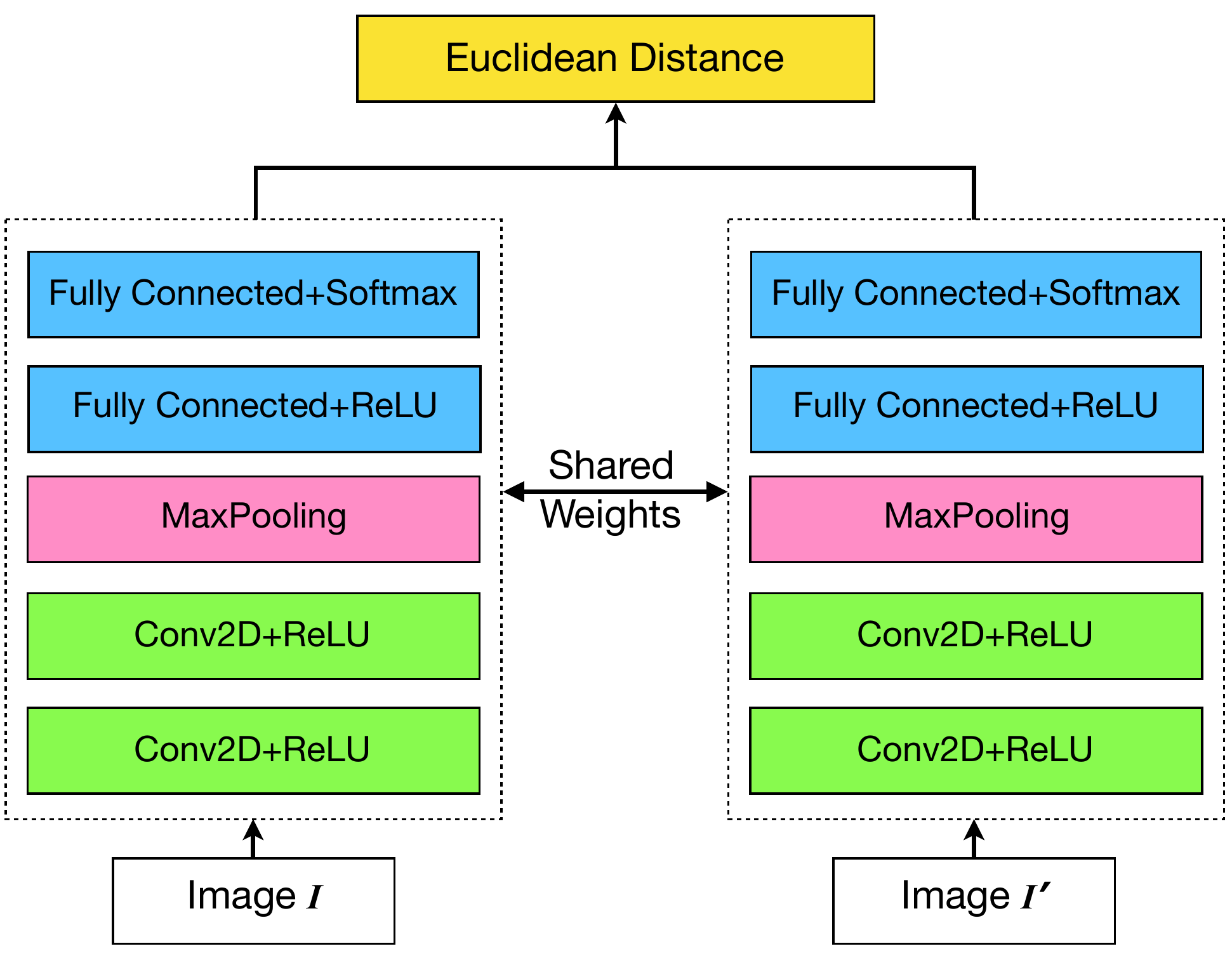}}
\caption{The architecture of a Siamese network which is used as our AE detector.}\label{fig:sianet}
\end{figure}

Given an input image $I$, 
when pre-processing is adopted, the input image $I$ and the pre-processed one $I^\prime$
may be very different even if $I$ is benign. On the other hand, the discrepancy
between \emph{the two images}, $I$ and $I^\prime$, may not be simply described using a single value and compared
with a threshold, as adopted by \emph{feature squeezing}~\cite{xu2017feature}. 
These are the main challenges in devising an accurate detection technique.

\begin{figure}[!t]
\centering
\centerline{\includegraphics[scale=0.45, trim=0 0 5 0,clip]{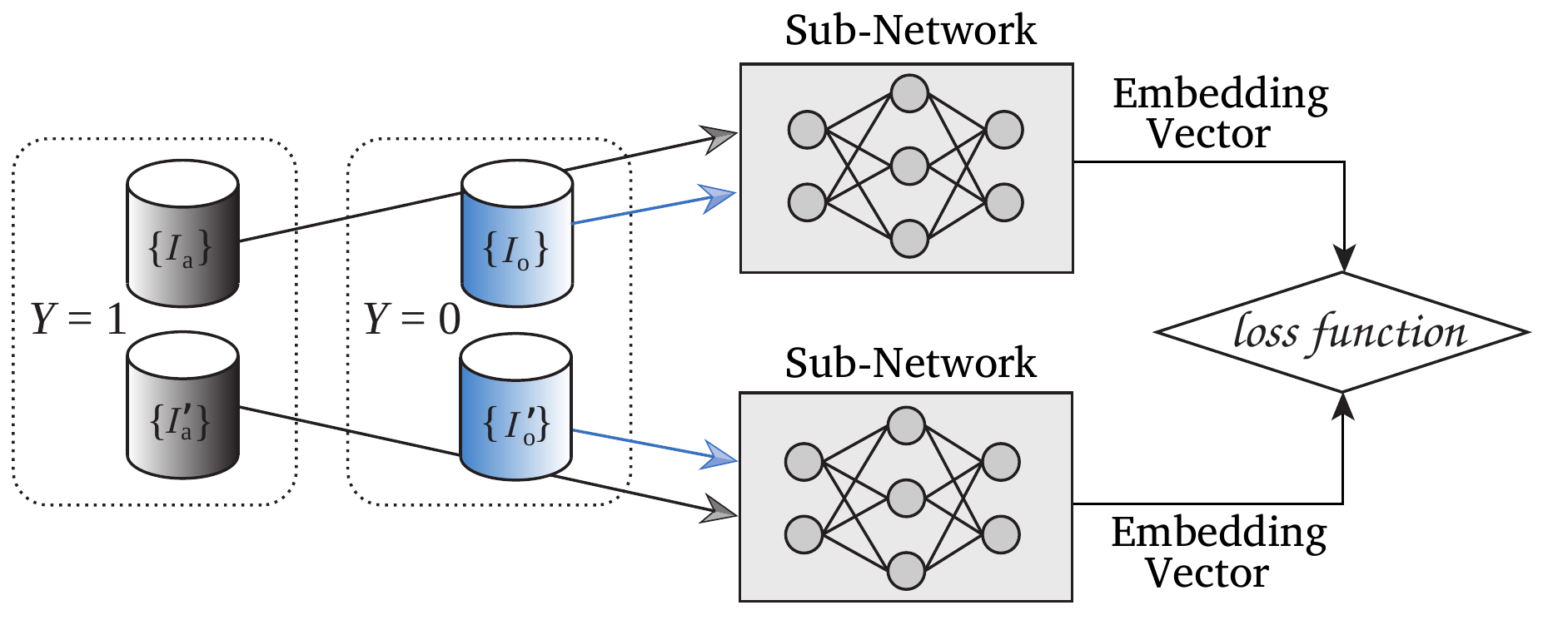}}
\caption{Training phrase of the AE detector based on a Siamese network.}\label{fig:train}
\end{figure}

We propose a Siamese-based $L_0$ AE detector with the help of 
a \emph{pre-processor}, which converts the AE detection problem into an image
comparison problem. Once the model with fine tuned weights is established (via training), 
the discrepancy between $I$ and $I^\prime$ can be extracted by the Siamese network. 
Taking the discrepancies as features, the model can predict whether the input image is adversarial or not. 

Figure~\ref{fig:sianet} illustrates the architecture of our Siamese network-based AE detector. In particular, we learn from the classical AlexNet~\cite{krizhevsky2012imagenet} to design our CNN-based sub-networks but only use a shallow network. The purpose is to explore how well the AE detector performs even when it only uses a simple network design. The details of the sub-network employed by each twin in the Siamese network are as follows:
\begin{small}
\begin{align*}
\mathit{CNN:}\ &\to conv(3\times3, 64)\to ReLU\\
          &\to conv(3\times3, 64)\to ReLU\\
          &\to maxpool(2\times2)\to dropout(0.3)\\
          &\to Flatten\\
          &\to linear(\_\_, 128) \to ReLU\to dropout(0.5)\\
          &\to linear(128,10) \to softmax.
\end{align*}
\end{small}Figure~\ref{fig:train} elaborates the training phrase of the proposed detector based on a Siamese network. 
Given an image $I$ and its pre-processed version $I^\prime$, 
    the Siamese network takes $\langle I, I^\prime \rangle$ as inputs, where the label is $0$ if $I$ is not an AE (denoted as $I_o$), or $1$ if $I$ is an AE (denoted as $I_a$).
Although it is difficult to use a formula to describe the discrepancy between the input pair $\langle I_a, I_a^\prime \rangle$ and
the consistency between $\langle I_o, I_o^\prime \rangle$, the Siamese network is effective in learning such relationship. 
Moreover, the consistency and discrepancy can be learned even when
a non-powerful pre-processor is adopted, such as a bit depth reducer (see Section~\ref{sec:var_pro}). 
The result of the last layer of each of the two sub-networks is fed to a contrastive loss function~\cite{hadsell2006dimensionality}:
$$(1-Y)\frac{1}{2}(D_W)^2+(Y)\frac{1}{2}\{(\max(0,m-D_W)\}^2$$
where $D_W$ is defined as the Euclidean distance between the outputs of the two sub-networks, $Y$ is a binary label assigned to the input pair, and $m>0$ is a margin used to define a radius around the output of one of the sub-networks. Finally, once the model is successfully trained, the Siamese network can be used to determine whether $I$ is an AE. 

Our evaluation shows that, even with a relatively small training
dataset and a network with very few layers, 
our detector can still achieve a very high accuracy. 

%% file: alg.tex
\begin{algorithm}[htbp] 
{\small
\caption{The Pre-processor based on Inpainting.} \label{alg:inpaint}
\KwIn{A color image $\mathcal{I}$; \\
      $\quad\quad\ \ \;\;$two bounds $\alpha$ and $\beta$ used to find extreme values;\\
      $\quad\quad\ \ \;\;$a parameter $\gamma$ to describe \textit{bright pixels} in natural images;\\ 
      $\quad\quad\ \ \;\;$a \textit{structuring element} $\mathcal{E}$ of the specified size and shape.} 
\KwOut{A processed color image, denoted by $\mathcal{S}$.}
\BlankLine
Normalize $\mathcal{I}\leftarrow [min(\mathcal{I})-\mathcal{I}]/[min(\mathcal{I})-max(\mathcal{I})]$\;
Extract three channels ($\mathcal{I}^R$, $\mathcal{I}^G$, $\mathcal{I}^B$) from $\mathcal{I}$\;
Initialize the masks $\mathcal{M}^R,\mathcal{M}^G,\mathcal{M}^B\leftarrow \{0\}$\;
\For {$\mathrm{each\ pixel}\ \mathcal{I}_i\in\mathcal{I}$,} 
{
\If{$(\mathcal{I}^R_i<\alpha)\vee[(\mathcal{I}^R_i>\beta) \wedge (\mathcal{I}^G_i\leq\gamma\ \vee\ \mathcal{I}^B_i\leq\gamma)]$}
{
$ \mathcal{M}^R_i \leftarrow 1 $\;
}
\If{$(\mathcal{I}^G_i<\alpha)\vee[(\mathcal{I}^G_i>\beta) \wedge (\mathcal{I}^R_i\leq\gamma\ \vee\ \mathcal{I}^B_i\leq\gamma)]$}
{
$ \mathcal{M}^G_i \leftarrow 1 $\;
}
\If{$(\mathcal{I}^B_i<\alpha)\vee[(\mathcal{I}^B_i>\beta) \wedge (\mathcal{I}^G_i\leq\gamma\ \vee\ \mathcal{I}^R_i\leq\gamma)]$}
{
$ \mathcal{M}^B_i \leftarrow 1 $\;
}
}
\For{$\mathrm{each\ pixel}\ \mathcal{M}^{\chi}_i\in \mathcal{M}^{\chi}, \mathrm{where}\ \chi:= R, G, B$, }{

\If{$\ \exists\ \mathit{N}(\mathcal{M}^{\chi}_i)>\mathcal{E},  \mathrm{\ s.t.\ } \mathcal{M}^{\chi}_j=1\ \wedge\ \mathcal{M}^{\chi}_j\in\mathit{N}(\mathcal{M}^{\chi}_i)$
}{$\mathcal{M}^{\chi}_j\leftarrow0$\;}
}
\For{ $\mathrm{each}\ \mathcal{I}^{\chi}:= \mathcal{I}^R, \mathcal{I}^G, \mathcal{I}^B$,}
{
$\mathcal{S}^{\chi}\leftarrow$Inpainting $\mathcal{I}^{\chi}$ according to $\mathcal{M}^{\chi}$\;
}
Reconstruct $\mathcal{S}$ with  $\mathcal{S}^R$, $\mathcal{S}^G$ and $\mathcal{S}^B$\;
\textbf{return} $\mathcal{S}$.
}
\end{algorithm}


%% file: eval.tex
\section{Evaluation}\label{sec:eval}

In this section, we evaluate our system on its detection and defense capability. We first 
describe the experimental settings and implementation (Section~\ref{sebsec:tar_model}) 
and discuss the datasets used in our evaluation (Section~\ref{sec:dataset}). 
We then evaluate the effect of our pre-processing method as a defense alone 
(Section~\ref{sec:eval_preproc}). 
Next we evaluate the accuracy of 
our system on detecting AEs generated by JSMA and CW-$L_0$  (Section~\ref{sec:effectiveness}), 
and the efficiency in terms of training and testing (Section~\ref{sec:efficiency}).
The resilience of our system against an adaptive attack is presented in Section~\ref{sec:adapt}.

It is worth noting that our proposed method can not only detect adversarial examples but also rectify the classification results. Thus, Section~\ref{sec:eval_preproc} shows that our pre-processor as a defense can \emph{individually} and functionally rectify the classification results of $L_0$ AEs. Section~\ref{sec:effectiveness} demonstrates that the detector (i.e., pre-processor plus the Siamese architecture) can distinguish AEs from benign images.

\subsection{Experimental Settings}\label{sebsec:tar_model}

\noindent \textbf{Threat model.}
We assume attackers have full knowledge on a trained target image classification model, 
but no ability to 
influence that model. Thus, given a trained target model, attackers can use the 
$L_0$ attacks including JSMA and CW-$L_0$ to generate AEs that will be 
misclassified by the target model. 

\vspace{3pt}
\noindent \textbf{Target models.} We use two popular datasets for the image classification task: MNIST and CIFAR-10. For each dataset, we build up two individual models for the two types of $L_0$ attacks.
Specifically, for MNIST, we set up a CNN-based classifier~\cite{yash2017apply} for JSMA, and reuse the model structure provided in~\cite{carlini2017towards}---which we denote as Carlini$_M$---for CW-$L_0$. 
For CIFAR-10, we select the 32-layered ResNet model based on a residual learning framework~\cite{he2016deep} for JSMA, and reuse the model structure given in~\cite{carlini2017towards}---which we denote as Carlini$_C$---for CW-$L_0$.
All the target models are trained from scratch. 

Table~\ref{tab:model} summarizes the classification 
accuracy on the testing data of each model. The accuracy of Carlini$_M$ and the CNN target model for MNIST is 99.26\% and 99.52\%, respectively; 
and the accuracy of Carlini$_C$ and the ResNet model for CIFAR-10 is 78.86\% and 91.96\%, respectively. Note that only those images which can \emph{be correctly classified} by the corresponding target models are used to generate AEs in the following experiments. 

\begin{table}[!t]\small
\centering
\renewcommand\arraystretch{1.02}
\begin{tabular}{c|l|c}
\specialrule{.1em}{.05em}{.05em} 
\textbf{\ \ \ Dataset\ \ \ }  & \textbf{Target Model}  & \textbf{Accuracy}   \\ \specialrule{.1em}{.05em}{.05em}
\multirow{2}{1.5cm}{\centering MNIST} &  Carlini$_M$~\cite{carlini2017towards}    &   99.26\%    \\ \cline{2-3} 
                           &  CNN~\cite{yash2017apply}  &   99.52\%    \\ \hline
\multirow{2}{1.5cm}{\centering CIFAR-10} & Carlini$_C$~\cite{carlini2017towards} &   78.86\%     \\ \cline{2-3} 
                              & ResNet~\cite{he2016deep}  &   91.96\%  \\ 

\specialrule{.1em}{.05em}{.05em} 
\end{tabular}
\caption{Classification accuracy of the target models.}\label{tab:model}
\end{table}

\vspace{3pt}
\noindent \textbf{Attacks.} 
For the target models Carlini$_M$ and Carlini$_C$, we reuse the code provided in~\cite{carlini2017towards} to generate CW-$L_0$ AEs.
The default parameters settings suggested by Carlini and Wagner~\cite{carlini2017towards} are as follows: the number of maximum iterations is 1000, the initial constant is 0.001, and the largest constant is $2^6$. 
To compare with the state-of-the-art works~\cite{xu2017feature,ma2019nic}, we follow these parameters settings.
Furthermore, for the target CNN and ResNet model, we generate AEs with JSMA by leveraging the Adversarial Robustness Toolbox (ART)~\cite{art2018}. 
We used the same parameters settings as~\cite{xu2017feature, ma2019nic}, i.e., $\theta=1, \gamma=0.1$. As both JSMA and CW-$L_0$ are targeted attacks, 
we designate the \emph{next} class as the target class.

Table~\ref{tab:attacks} reports the results of the AEs. The \emph{success rate} 
is defined as the probability that an adversary achieves their goal. 
For a targeted attack, it is only considered 
a success if the model predicts the target class.
Note that we only use the AEs that can \emph{successfully attack the target models} 
to evaluate the performance of our system on detecting AEs. 

\input{table2_4.tex}
\input{table5_6.tex}
\noindent \textbf{Implementation.}
We implement our Siamese-based detector 
in Python using the Keras~\cite{chollet2015keras} 
platform with TensorFlow~\cite{abadi2016tensorflow} as backend. Keras provides a large 
number of high-level neural network APIs and can run on top of TensorFlow. The Telea's inpainting algorithm~\cite{telea2004image} is  
implemented based on Open Source Computer Vision Library (OpenCV)~\cite{bradski2017opencv}. 

The experiments were performed on a computer running the Ubuntu 18.04 operating system 
with a 64-bit 3.6 GHz Intel\textsuperscript{\textregistered} Core\textsuperscript{(TM)} 
i7 CPU, 16 GB RAM and GeForce GTX 1070 GPU.

\subsection{Data Preparation} \label{sec:dataset}
We generate AEs based on two image datasets, i.e., CIFAR-10 and MNIST.

\vspace{3pt}
\noindent \textbf{CIFAR-10} contains 60,000 color images; each is assigned to one of ten different classes, 
such as dog, frog and ship. CIFAR-10 is split into the training and testing dataset, which 
contains 50,000 and 10,000 images, respectively. 

We first filter out those images that cannot be correctly classified by the corresponding target model.
We then use the CW-$L_0$ algorithm to generate AEs that can \emph{successfully attack}
the Carlini$_C$ model~\cite{carlini2017towards}, and create two dis-joint datasets, denoted as $\mathcal{D_C}$\texttt{-CWL0-Train} and $\mathcal{D_C}$\texttt{-CWL0-Test}. 
In detail, $\mathcal{D_C}$\texttt{-CWL0-Train} contains 10,000 legitimate images and 
10,000 AEs. $\mathcal{D_C}$\texttt{-CWL0-Test} contains 1,000 benign images and 1,000 AEs. Next, we follow the similar method on CIFAR-10 but instead using JSMA to generate AEs based on ResNet classifier~\cite{he2016deep}. 
As a result, we obtain two dis-joint datasets, denoted as $\mathcal{D_C}$\texttt{-JSMA-Train} and $\mathcal{D_C}$\texttt{-JSMA-Test}.
There are 10,000 legitimate images and 10,000 AEs in training set. 
There are 1,000 legitimate images and 1,000 AEs in testing set. 

\vspace{3pt}
\noindent \textbf{MNIST} contains 70,000 8-bit grayscale images of hand-written digits. Each image is 
assigned a label from 0 to 9. MNIST is split into the training and testing dataset, 
which contains 60,000 and 10,000 images, respectively. We carry out similar procedures on MNIST to create a training and testing set but using different target models.
As a result, we have $\mathcal{D_M}$\texttt{-CWL0-Train} and $\mathcal{D_M}$\texttt{-CWL0-Test} based on Carlini$_M$ model~\cite{carlini2017towards},
as well as $\mathcal{D_M}$\texttt{-JSMA-Train} and $\mathcal{D_M}$\texttt{-JSMA-Test} based on CNN~\cite{yash2017apply} model. The sizes of these datasets are the same as their counterparts in CIFAR-10. Considering that CIFAR-10 is a more challenging dataset compared with MNIST, we will spend more space on explaining the results for CIFAR-10 in the following experiments.

\vspace{3pt}
\noindent\textbf{\emph{Note}} that all the aforementioned legitimate images can be classified correctly by the target model, and all the AEs can successfully fool the corresponding target model.

\subsection{Effectiveness of Pre-processor as Defense}\label{sec:eval_preproc}

To mislead a classifier to predict a specific target class, the adversarial 
perturbations produced by an $L_0$ attack such as JSMA or CW-$L_0$ are introduced 
intentionally instead of randomly. Moreover, the adversarial strength of an 
$L_0$ attack limits the number of pixels that can be manipulated; and as a 
result, the manipulated pixels need to have significant changes. 
The proposed inpainting-based pre-processor is to eliminate the possible adversarial 
pixels while preserving the benign ones. Therefore, the inpainting-based pre-processor can also be considered as a defense against $L_0$ attacks. 

\vspace{3pt}
\noindent\textbf{Inpainting-based pre-processor for color images.} 
We first evaluate the effectiveness of the inpainting-based pre-processor as a defense against $L_0$ attack on CIFAR-10. 
The inpainting-based algorithm has two parameters: the threshold $\alpha$ extracts the pixels whose values tend to be very small, and $\beta$ is used to screen all pixels whose values tend to be extremely large. 
We use 1,000 AEs in $\mathcal{D_C}$\texttt{-CWL0-Test}
to evaluate the effectiveness of the inpainting-based pre-processor as a defense against CW-$L_0$ attacks
and examine its performance with varying values of $\alpha$ and $\beta$. 
Without pre-processing, these AEs result in 0\% classification accuracy when using the Carlini$_C$ model~\cite{carlini2017towards}. After pre-processing, each recovered AE is analyzed by the model to predict 
a class label. Table~\ref{preproc_cw0} shows the results. When $\alpha = 0.1$ and $\beta=0.7$, the performance is the best---the 
classification accuracy on these AEs is increased from 0\% to 87.3\%. 

We then use 1,000 AEs in $\mathcal{D_C}$\texttt{-JSMA-Test} to evaluate the effectiveness of the inpainting-based pre-processor as
a defense against JSMA attack. 
Without pre-processing, these AEs result in 0\% classification accuracy of 
the ResNet model~\cite{he2016deep}. After applying inpainting-based pre-processor to rectify those AEs, each recovered AE is analyzed by the ResNet model to predict 
a class label. Table~\ref{preproc_jsma} shows the results. We can 
see that when $\alpha = 0.0$ and $\beta=0.8$, the performance is the best---the 
classification accuracy on these AEs is increased from 0\% to 96.1\%. 
This classification accuracy is higher than 87.3\% given by the previous Carlini$_C$ model. 
Moreover, the ResNet model is more robust against the benign perturbations introduced by the inpainting procedure.

As a comparison, we examine the impact of both SVD compression and median filter on AEs generated by $L_0$ attacks. First, as a low-pass filter, SVD compression is usually used to reduce noise in images. 
As shown in Table~\ref{svd}, when varying the loss ratio of SVD compression, the classification accuracy on the processed AEs is very low---at most 44.5\% and 27.4\% for CW-$L_0$ AEs and JSMA AEs, respectively. 
Note that we only use those images which can be correctly classified by the target model to generate AEs; thus the maximum classification accuracy given by the target model here is 100\%.
Therefore, the experiment suggests that the perceptible perturbations introduced by an $L_0$ attack are very difficult to be reduced when only using the frequency domain filters.
Alternatively,~\cite{xu2017feature} and~\cite{guo2017countering} claim that median filter is particularly effective in mitigating adversarial examples generated by an $L_0$ attack because such perturbations are very similar to salt-and-pepper noises. In our experiments, after applying the median filter to process AEs in $\mathcal{D_C}$\texttt{-CWL0-Test} and $\mathcal{D_C}$\texttt{-JSMA-Test}, 
the classification accuracy given by Carlini$_C$ and ResNet is 79.8\% and 85.3\%, respectively; both are lower than the proposed defense.

\vspace{3pt}
\noindent\textbf{Inpainting-based pre-processor for gray images.} We can observe similar results on MNIST when taking advantage of the inpainting-based pre-processor as a defense against an $L_0$ attack. 
Our experiment shows that processing the 1,000 AEs in $\mathcal{D_M}$\texttt{-CWL0-Test} with the proposed inpainting-based method results in a significant increase of the classification accuracy on the Carlini$_M$ model~\cite{carlini2017towards}---from 0\% to 88.2\%. 
Similarly, after using the proposed inpainting-based method on the 1,000 AEs in $\mathcal{D_M}$\texttt{-JSMA-Test}, the recovered AEs result in the classification accuracy on the CNN model~\cite{yash2017apply} to 
increase from 0\% to 86.1\%. All of the results above are obtained when $\alpha=0.1$ and $\beta=0.8$. When varying the value of $\alpha$ from 0.1 to 0.2 and the value of $\beta$ from 0.7 to 0.8, the classification accuracy increases slowly (84.9\% at least).

As a comparison, Bafna et al.~\cite{bafna2018thwarting} independently focus on the $L_0$ attacks, and propose a defense based on Fourier 
transform. But our approach outperforms theirs---after applying their defense algorithm against CW-$L_0$, their classification 
accuracy on the MNIST testing set is only
72.8\%. Note that they did not conduct experiments on standard color-image datasets such as CIFAR-10. 

\vspace{3pt}
\noindent \textbf{Impact on benign images.}
To investigate the impact of the defense methods on benign images, we  
first carry out an experiment on the 1,000 benign images from $\mathcal{D_C}$\texttt{-JSMA-Test}. Specifically, we use 
the ResNet model~\cite{he2016deep} to classify each color image after the inpainting-based 
defense is applied. The classification accuracy on these processed images only 
decreases from 100\% to 95.6\%. Next, we conduct a similar experiment on the 1,000 benign images from $\mathcal{D_M}$\texttt{-JSMA-Test}. 
We use the CNN model~\cite{yash2017apply} to classify each gray image after the inpainting-based 
defense  is applied. As a result, the classification accuracy on these processed images only 
decreases from 100\% to 99.7\%. The results show that a very small impact is imposed 
on classifying benign images.

\vspace{3pt}
\noindent \textbf{\emph{Summary.}} 
Therefore, the proposed inpainting-based 
algorithm is effective in defending against $L_0$ attacks such as CW-$L_0$ and JSMA. Moreover, our defense methods
have a very small impact on benign images, which implies it can be directly applied
without relying on detection.

\subsection{Detecting $L_0$ Adversarial Inputs} \label{sec:effectiveness}

We next evaluate the effectiveness of our system on detecting AEs generated by $L_0$ attacks. 

\subsubsection{Detection Efficacy}
We evaluate the detection performance of the proposed scheme against CW-$L_0$ and JSMA attack. 
The inpainting-based pre-processor is used to create input pairs to the Siamese network.

\vspace{3pt}
\noindent \textbf{Color images.} 
The two training datasets, $\mathcal{D_C}$\texttt{-CWL0-Train} and $\mathcal{D_C}$\texttt{-JSMA-Train}, are used to train our system individually for 200 epochs 
using early stopping configured with a minimum accuracy change of 0.001 and 50 patience steps. If an accuracy change is less than 0.001, we consider that there is
no improvement of the model performance; after 50 epochs with no improvement, 
the training is stopped. We save the resulting models as the base models.

We now evaluate the detection accuracy of the base models against CW-$L_0$ and JSMA attack on $\mathcal{D_C}$\texttt{-CWL0-Test} and $\mathcal{D_C}$\texttt{-JSMA-Test}, respectively.
Each dataset contains 1,000 benign images and 1,000 AEs. 
We plot the ROC (receiver operating characteristic) curves, which are showed in Figure~\ref{fig:roc_detector}(a). 
We can achieve the AUC values of 98.69\% and 99.94\% for the two $L_0$ attacks. 
Table~\ref{tab:detector} shows more detailed results evaluated on the testing set.

We consider the adversarial images as positive, and the benign images as negative. 
Thus, the recall value (i.e., the detection rate) is
the ratio of the number of successfully detected AEs to the total 
number of AEs; 
and False Positive Rate (FPR) is the fraction of the negative testing data (i.e., benign images) that is misclassified as positive. 
In practice, the distribution of adversarial and benign images are not balanced---most of the images should be benign. Thus, FPR is a very important metric to evaluate the model performance; a lower FPR indicates that the system makes fewer mistakes for benign images.

\input{fig4_table7.tex}

As shown in Table~\ref{tab:detector}, when analyzing AEs generated by the CW-$L_0$ attack, the detection rate of $\mathcal{D_C}$\texttt{-CWL0-Test} is 97.1\% and the FPR is 5.5\%.
When analyzing AEs generated by the JSMA attack, the detection rate of $\mathcal{D_C}$\texttt{-JSMA-Test} is 99.7\% and the FPR is as low as 0.0\%. 

\vspace{3pt}
\noindent \textbf{Gray images.} We follow the same configurations to conduct an experiment on MNIST. Because gray images only have one channel, training a Siamese network on MNIST is simpler than that on CIFAR-10. We thus only train the detector for 100 epochs using an early stopping with 30 patience steps. 

We evaluate the detection accuracy of the base models against CW-$L_0$ and JSMA attacks on $\mathcal{D_M}$\texttt{-CWL0-Test} and $\mathcal{D_M}$\texttt{-JSMA-Test}, respectively.
We plot the ROC curves as the Figure~\ref{fig:roc_detector} (b) shows. 
The AUC value can achieve 99.84\% and 99.93\%. In Table~\ref{tab:detector}, we can observe similar results as the experiments given on CIFAR-10. When facing CW-$L_0$ attacks, the detection rate for the AEs from $\mathcal{D_M}$\texttt{-CWL0-Test} is 99.5\%, and the FPR is 0.7\%. When facing JSMA attacks, the detection rate for the AEs from $\mathcal{D_M}$\texttt{-JSMA-Test} can achieve
99.7\%, and the FPR is as low as 0.1\%.

\vspace{3pt}
\noindent \textbf{\emph{Comparison.}} 
We compare the proposed system with the state-of-the-art AE detectors, including feature squeezing~\cite{xu2017feature} and NIC~\cite{ma2019nic};
both of them show that their systems are able to effectively detect $L_0$ AEs. Moreover, feature squeezing~\cite{xu2017feature} uses multiple feature squeezers, and we only compare our system with the \emph{best} results of their work.
To this end, we train two comprehensive models for color images and gray images, respectively. 
Specifically, for color images, we train the detector using both $\mathcal{D_C}$\texttt{-CWL0-Train} and $\mathcal{D_C}$\texttt{-JSMA-Train}. 
For gray images, we train another detector using $\mathcal{D_M}$\texttt{-CWL0-Train} and $\mathcal{D_M}$\texttt{-JSMA-Train}.
We summarize the adversarial detection rate and FPR in Table~\ref{tab:compare}. 
For CIFAR-10, the detection rate of feature squeezing~\cite{xu2017feature} on CW-$L_0$ and JSMA attacks is 98.1\% and 83.7\%, respectively, and its FPR---the percentage of the benign images among all the testing benign images that is misclassified as positive---is 4.9\%. 
NIC~\cite{ma2019nic} can achieve the detection rate of 98.0\% and 94.0\% on CW-$L_0$ and JSMA attacks respectively, and its FPR is 3.8\%. 
Our \textsc{AEPecker} outperforms theirs---we can achieve the detection rate of 98.4\% and 99.5\% for the two types of $L_0$ attacks and our FPR is only 2.0\%. 
With repect to MNIST, the detection rate of our model is comparable with feature squeezing~\cite{xu2017feature} 
and NIC~\cite{ma2019nic}. Moreover, \textsc{AEPecker} achieves the lowest FPR for both CIFAR-10 and MNIST. Therefore, our proposed detector outperforms the  two state-of-the-art detectors.

\subsubsection{Transferability} \label{subsec:eval-transfer}

This experiment is to evaluate the transferability of our system: whether our system trained on one type of $L_0$ AEs can be directly applied to detect another type of $L_0$ AEs that have not been seen during training without any adaptation. 
To this end, we train our system using $\mathcal{D_C}$\texttt{-JSMA-Train} and use $\mathcal{D_C}$\texttt{-CWL0-Test} to test the detector. 
The result shows that the detection rate 
is as high as 99.4\%. Similarly, 
we train our system using $\mathcal{D_C}$\texttt{-CWL0-Train} and use $\mathcal{D_C}$\texttt{-JSMA-Test} to test the detector. The result shows that the detection rate is as high as 98.7\%. 

The similar results are obtained for MNIST: if we use $\mathcal{D_M}$\texttt{-JSMA-Train} to train the Siamese network and use $\mathcal{D_M}$\texttt{-CWL0-Test} to test the detector, the detection rate 
is 96.3\%; if our system is tained using $\mathcal{D_M}$\texttt{-CWL0-Train} and tested on $\mathcal{D_M}$\texttt{-JSMA-Test}, the detection rate can achieve 95.4\%. 

\vspace{3pt}
\noindent \textbf{\emph{Summary.}} 
Therefore, our system has good transferability; our system trained on AEs generated by one $L_0$ attack can be directly applied to detect AEs generated by another $L_0$ attack without any adaptation.


\subsubsection{Pre-processor Study} \label{sec:var_pro} 

We next conduct an experiment to examine the impact of the pre-processor; specifically, we would like to see what the detection accuracy will be if a weak pre-processor is adopted. The \emph{weak} here means the manipulated AEs through such a pre-processor still cannot be classified correctly by the target model with a high possibility. Through this, we will show that even with a weak pre-process, our system can still achieve a high detection accuracy---this means that a perfect pre-processor is unnecessary for our Siamese-based detector to achieve a high success rate of detection.

\begin{figure}[!t]
\centering
\centerline{\includegraphics[scale=0.22]{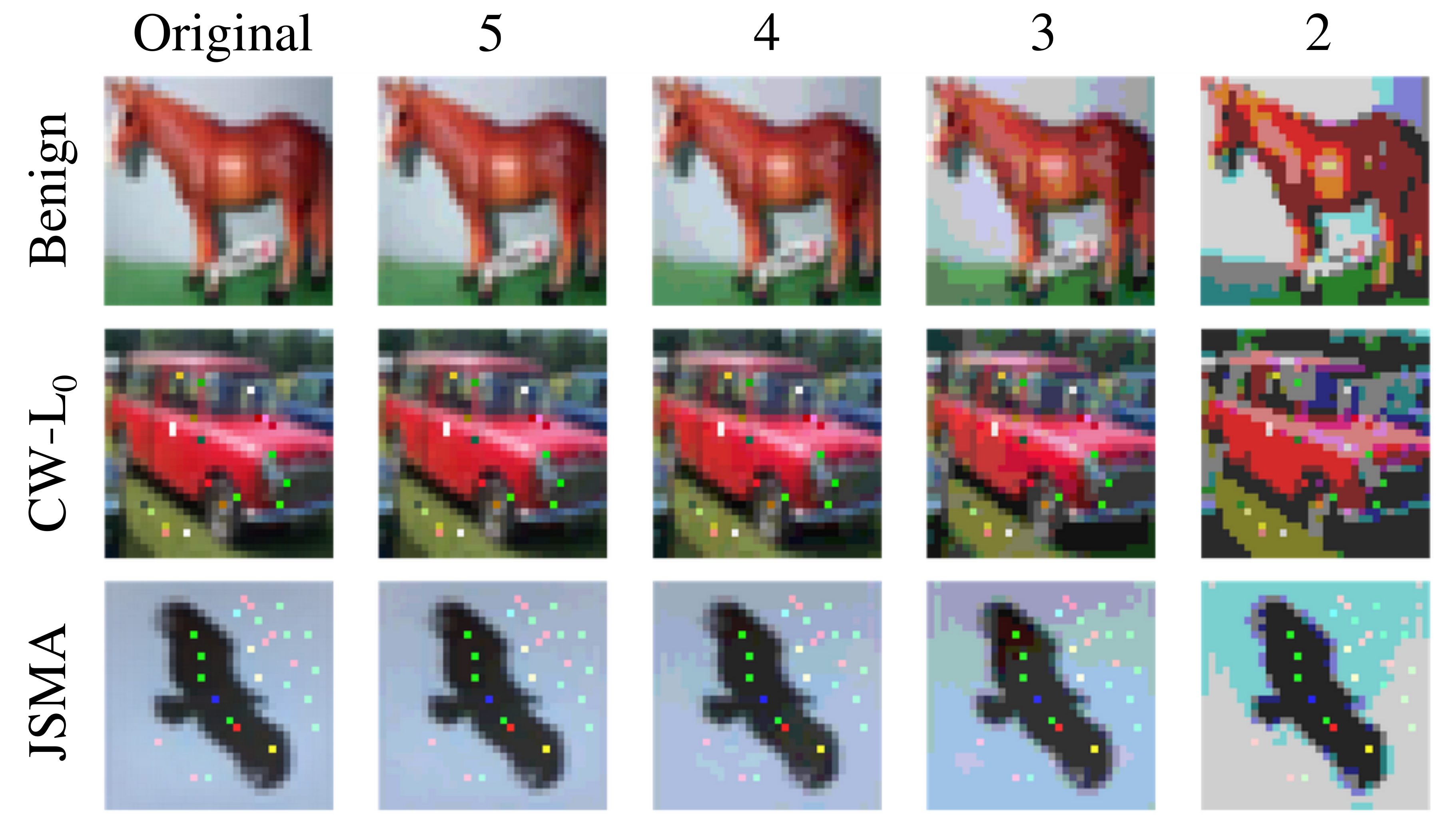}}
\caption{Image examples from CIFAR-10 after applying bit depth reduction. Given the different numbers of bit depth, the first row displays a benign image and its processed versions; 
the second row displays an AE generated by CW-$L_0$ and its corresponding processed images;
the third row displays an AE generated by JSMA and its corresponding processed images.}\label{fig:bit_depth}
\end{figure}

Without loss of generality, we use color images as an example for the following discussion. For color images such as CIFAR-10, each channel of RGB is encoded by 8 bits. As Figure~\ref{fig:bit_depth} shows, we can reduce the original 8 bits to fewer bits without influencing the image recognizability for human eyes. Figure~\ref{fig:bit_depth} also shows that it is very difficult to remove those striking adversarial perturbations introduced by $L_0$ attacks only with such an approach. 
Moreover, the original $L_0$ AEs can mislead the target neural networks to a classification accuracy of 0\%.
After applying bit depth reduction, the classification accuracy for AEs in the testing datasets is calculated.
The experiment results are shown in Table~\ref{fig:bit_reduce}, which suggest that processing the AEs generated by JSMA and CW-$L_0$ with bit depth reduction cannot increase the classification accuracy of the target model. 
Therefore, the bit depth reduction approach only has a very limited capability to defend against $L_0$ attacks. 

We thus choose the bit depth reduction as a weak pre-processor for color images. The experiment results show that even when a weak pre-processor (such as bit depth reduction) is applied, our Siamese-based detector still can achieve a very good performance. The detection rates for AEs generated by JSMA and CW-$L_0$ are 99.6\% and 99.4\%, respectively; and the FPR is 2.1\%. Xu et al. also use bit depth reduction as a pre-processor~\cite{xu2017feature}; however, the best detection rates provided by their system for AEs generated by JSMA and CW-$L_0$ are 4.1\% and 36.5\% respectively, and its FPR is 5\%. 

\vspace{3pt}
\noindent \textbf{\emph{Summary.}} 
Therefore, our proposed Siamese-based detector outperforms the state-of-the-art method when using the same weak pre-processor. The result also demonstrates that the good performance of our detector does not rely on a perfect pre-processor, but is due to the Siamese network design.

\subsection{Efficiency} \label{sec:efficiency}

\noindent \textbf{Training time.}
It is widely known that neural networks usually require a large amount of data
and time for training. However, as our sub-networks employed within the Siamese 
architecture are quite simple and shallow, the training is very efficient.
For example, for $\mathcal{D_M}$\texttt{-JSMA-Train} and $\mathcal{D_C}$\texttt{-JSMA-Train}, each epoch with 20,000 
images (10,000 benign images and 10,000 AEs) only takes 5 and 7 seconds, respectively. On the
other hand, due to the simple and shallow sub-networks, with a relatively 
small training set, our Siamese neural-network-based \textsc{AEPecker} can still achieve high detection accuracies 
(Section~\ref{sec:effectiveness}). 
Moreover, the training time is linear with respect to the number of epochs and the number of training samples for each epoch. 

Our experiment results show that our system trained on CIFAR-10 
and MNIST can converge very quickly and achieve 
high accuracy within 100 and 200 epochs, respectively---thus, the training only 
requires around 8 minutes and 23 minutes, respectively. 

\begin{table}[!t]\small
\centering
\renewcommand\arraystretch{1.02}
\begin{tabular}{c||c|c|c|c}
\hline
\multirow{2}*{\textbf{Datasets}}  & \multicolumn{4}{c}{\textbf{Bit Depth}}   \\ \cline{2-5}
                 &  2-bit & 3-bit & 4-bit & 5-bit \\ \hline 
    $\mathcal{D_C}$\texttt{-JSMA-Test} &22.2\% & 27.1\%  & 21.2\% & 12.0\% \\ \hline 
    $\mathcal{D_C}$\texttt{-CWL0-Test} &51.2\% & 56.6\%  & 55.1\% & 51.5\% \\ \hline 
\end{tabular}
\caption{The classification accuracy for AEs in testing datasets
after applying bit depth reduction.}
\label{fig:bit_reduce}
\end{table}

\vspace{3pt}
\noindent \textbf{Testing time.}
The trained detector can detect an AE very fast. For example, \textsc{AEPecker} only takes approximately 0.5ms on average to detect whether an image from CIFAR-10 is adversarial or not. 

%% file: table2_4.tex
\begin{table*}
\begin{minipage}{0.33\textwidth}
\small \vspace{-20pt}
\centering
\renewcommand\arraystretch{1.02}
\begin{tabular}{c|c|c}
\specialrule{.1em}{.05em}{.05em} 
\textbf{\ \ Dataset\ \ }  & \textbf{Attack}  & \textbf{Success rate}   \\ \specialrule{.1em}{.05em}{.05em}
\multirow{2}{1.5cm}{\centering MNIST} &  CW-$L_0$~\cite{carlini2017towards}  &  100\%  \\ \cline{2-3} 
                              &  JSMA~\cite{papernot2016limitations}     &   81.6\%  \\ \hline
\multirow{2}{1.5cm}{\centering CIFAR-10} &  CW-$L_0$~\cite{carlini2017towards}  &  100\%  \\ \cline{2-3} 
                              &  JSMA~\cite{papernot2016limitations}     &   99.8\%  \\ 
\specialrule{.1em}{.05em}{.05em} 
\end{tabular} 
\caption{Evaluation of the $L_0$ attacks}\label{tab:attacks}
\end{minipage}
\quad
\begin{minipage}{.31\textwidth}
\small
\centering
\renewcommand\arraystretch{1.02}
\begin{tabular}{|c|ccc|}
\hline
\diagbox{$\beta$}{$\alpha$} & 0.0     & 0.1     & 0.2 \\
\hline
0.6 & 81.3\% & 86.9\% & 84.2\% \\
0.7 & 80.5\% & 87.3\% & 86.5\% \\ 
0.8 & 76.0\% & 86.2\% & 86.5\% \\ 
\hline
\end{tabular}
\caption{The classification accuracy on AEs in $\mathcal{D_C}$\texttt{-CWL0} \texttt{-Test} after using inpainting-based pre-processors.}\label{preproc_cw0}
\end{minipage}
\quad
\begin{minipage}{.31\textwidth}
\small
\centering
\renewcommand\arraystretch{1.02}
\begin{tabular}{|c|ccc|}
\hline
\diagbox{$\beta$}{$\alpha$} & 0.0     & 0.1     & 0.2 \\
\hline
0.6 & 90.0\% & 81.2\% & 63.2\% \\
0.7 & 94.1\% & 88.8\% & 74.5\% \\ 
0.8 & 96.1\% & 91.2\% & 77.3\% \\ 
\hline
\end{tabular}
\caption{The classification accuracy on AEs in $\mathcal{D_C}$\texttt{-JSMA} \texttt{-Test} after using inpainting-based pre-processors.}\label{preproc_jsma}
\end{minipage}
\end{table*}

%% file: table5_6.tex
\begin{table*}
\centering
\hspace{3pt}
\begin{minipage}{0.35\textwidth}
\vspace{6pt}
\small
\centering
\renewcommand\arraystretch{1.02}
\begin{tabular}{c|ccc}
\specialrule{.1em}{.05em}{.05em} 
Loss ratio & 60\%    & 40\%    & 20\%  \\ \specialrule{.1em}{.05em}{.05em} 
JSMA  & 27.4\% & 17.9\% & 5.4\%  \\ 
CW-$L_0$ & 44.5\% & 35.1\% & 21.4\% \\ \specialrule{.1em}{.05em}{.05em} 
\end{tabular}
\caption{The classification accuracy on testing datasets after applying SVD compression.}\label{svd}
\end{minipage}
\quad
\begin{minipage}{.61\textwidth}
\small
\centering
\renewcommand\arraystretch{1.02}
\begin{tabular}{c|c|c|c|c|c|c}
\hline
\textbf{Dataset} & \textbf{Attack}   & \textbf{Accuracy} & \textbf{Precision} & \textbf{Recall}  & \textbf{F1 Score} & \textbf{FPR}  \\  \hline
\multirow{2}*{CIFAR-10}& JSMA & 99.85\%   & 100.0\%   & 99.70\% & 99.85\%   & 0.0\%   \\ \cline{2-7}
                         & CW-$L_0$ & 95.80\%   & 94.64\%  & 97.10\% & 95.85\%   & 5.5\%   \\ \hline
\multirow{2}*{MNIST} & JSMA & 99.80\%  &  99.90\%  & 99.70\% & 99.80\%   & 0.1\%  \\ \cline{2-7}
 & CW-$L_0$ & 99.40\%   & 99.30\%  & 99.50\% & 99.40\%   & 0.7\%   \\ \hline
\end{tabular}
\caption{The detection performance of the proposed system.}\label{tab:detector}
\end{minipage}
\end{table*}

%% file: fig4_table7.tex
\begin{figure*}[!t]
\begin{minipage}{.5\textwidth}
    \subfloat[CIFAR-10]{\includegraphics[scale=0.3,trim=10 0 0 0, clip]{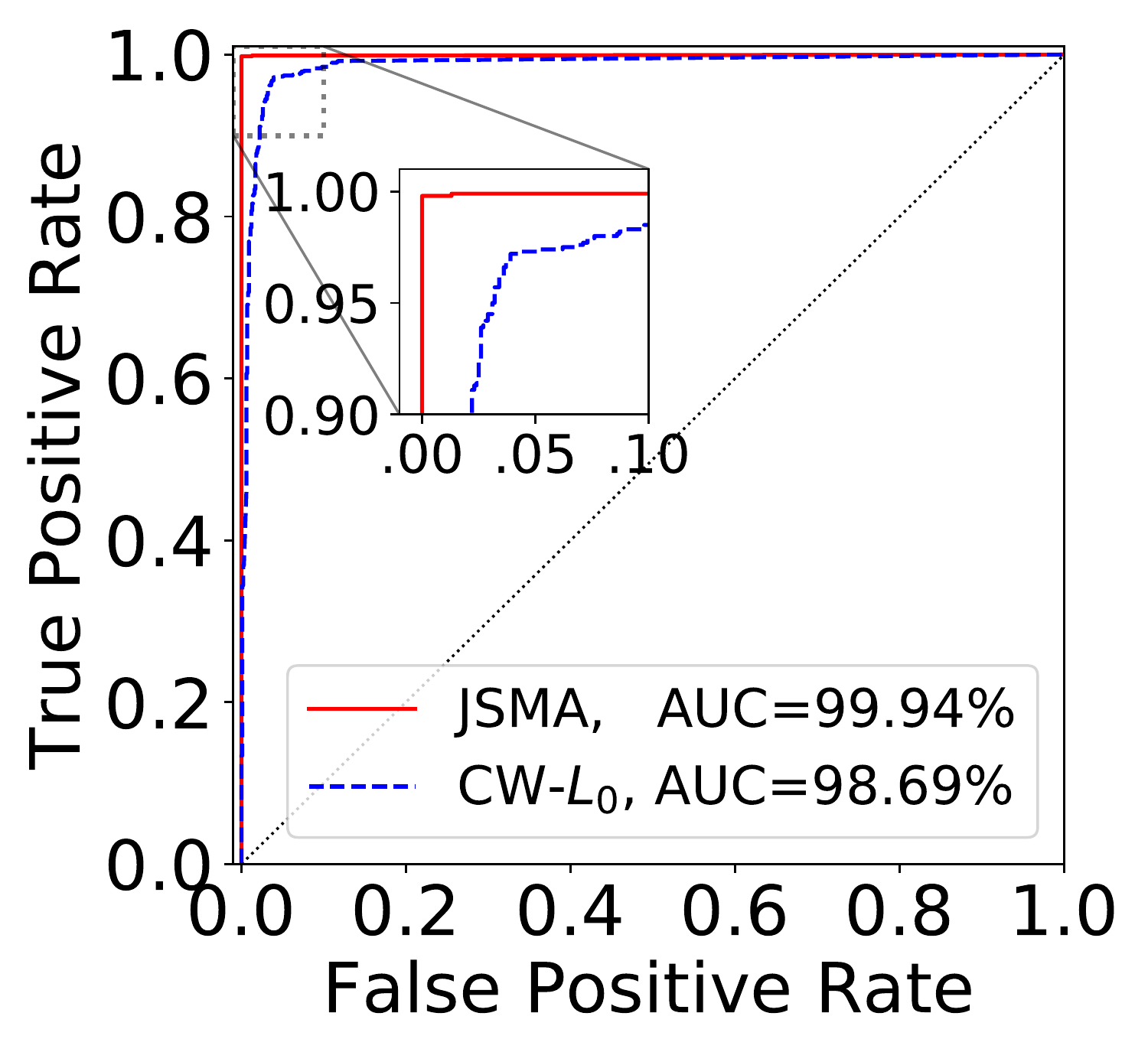}}
    \subfloat[MNIST]{\includegraphics[scale=0.3,trim=0 0 0 0, clip]{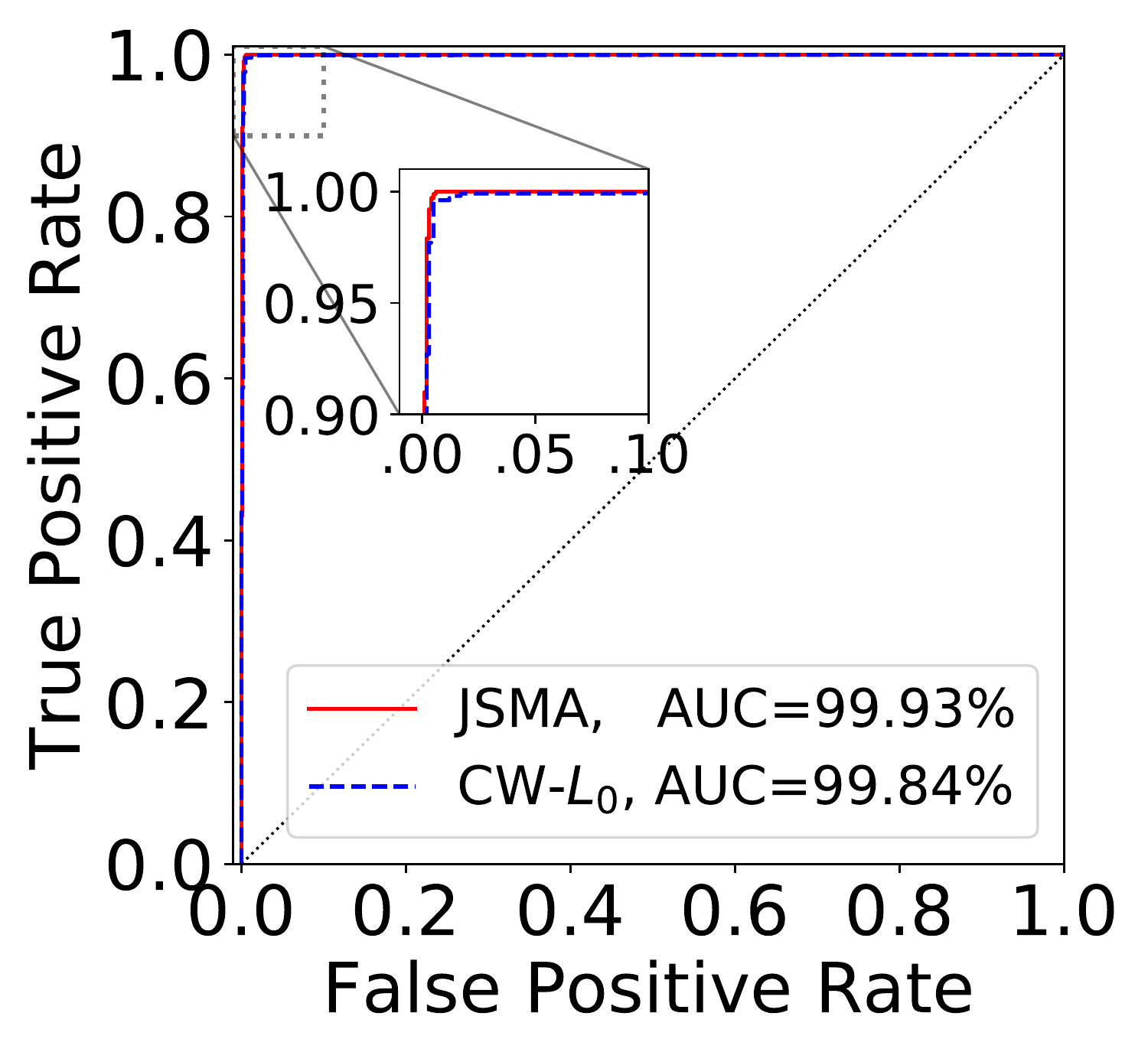}}     \setlength{\abovecaptionskip}{5pt}
    \caption{ROC curves for different datasets.} \label{fig:roc_detector}
\end{minipage}
\quad\ \ 
\begin{minipage}{.45\textwidth}
\vspace{16pt}
\centering
\captionsetup{type=table}
\setlength{\belowcaptionskip}{0pt}
\input{table7.tex}
\setlength{\abovecaptionskip}{28pt}
\captionof{table}{Comparison with state-of-the-art detectors in terms of FPR and detection rate.}\label{tab:compare} 
\end{minipage}

\end{figure*}

%% file: table7.tex
\centering
\renewcommand\arraystretch{1.02}
\begin{tabular}{c|c|c|c|c}
\specialrule{.1em}{.05em}{.05em}
      \textbf{Dataset}                    & \textbf{Detector} & \textbf{FPR}   & \textbf{CW-$L_0$} & \textbf{JSMA}   \\ \specialrule{.1em}{.05em}{.05em}
\multirow{3}{*}{CIFAR-10} & \textsc{AEPecker}       & 2.0\% & 98.4\%   & 99.5\% \\ \cline{2-5} 
                          & \textsc{ FS}~\cite{xu2017feature}       & 4.9\% & 98.1\%   & 83.7\% \\ \cline{2-5} 
                          & \textsc{NIC}~\cite{ma2019nic}     & 3.8\% & 98.0\%   & 94.0\% \\ \hline
\multirow{3}{*}{MNIST}    & \textsc{AEPecker}      & 0.4\% & 99.1\%   & 99.3\% \\ \cline{2-5} 
                          & \textsc{ FS}~\cite{xu2017feature}     & 4.0\% & 91.1\%   & 100\%  \\ \cline{2-5} 
                          & \textsc{NIC}~\cite{ma2019nic}      & 3.7\% & 100\%    & 100\%  \\ \specialrule{.1em}{.05em}{.05em}
\end{tabular}


%% file: adaptive.tex
\section{Adaptive Attack}\label{sec:adapt}
An adversary who knows the details of \textsc{AEPecker} will
try to adapt the attacks. Thus,
we seek to understand the resilience of \textsc{AEPecker} to adaptive 
attacks by answering the following questions.
(\textbf{Q1}) What is the percentage of the high-amplitude altered pixels in AEs
generated using non-adaptive $L_0$ attacks? Exploration of this question 
not only helps us understand $L_0$ attacks and why the proposed detection
and defense techniques work well, but also guides the adversary to adapt $L_0$ attacks.
(\textbf{Q2}) How difficult is it for an adversary to adaptively generate $L_0$ AEs that
bypass our detection?

To answer these questions, we launch an adaptive $L_0$ attack by adopting a similar method described in~\cite{he2017adversarial}, which has successfully demonstrated a capability to impede \textit{feature squeezing}~\cite{xu2017feature}. Our implementation is based on the source code given by~\cite{carlini2017towards}. To generate $L_0$ AEs, after each step of stochastic gradient descent (SGD), an intermediate distorted image is generated as a resolution of the optimizer.
Each time the optimizer runs, 
the process tries to minimizes the number of altered pixels and, in the meanwhile, keep the targeted attack successful. 

\begin{figure}[!t]
\centering
\centerline{\includegraphics[scale=0.5]{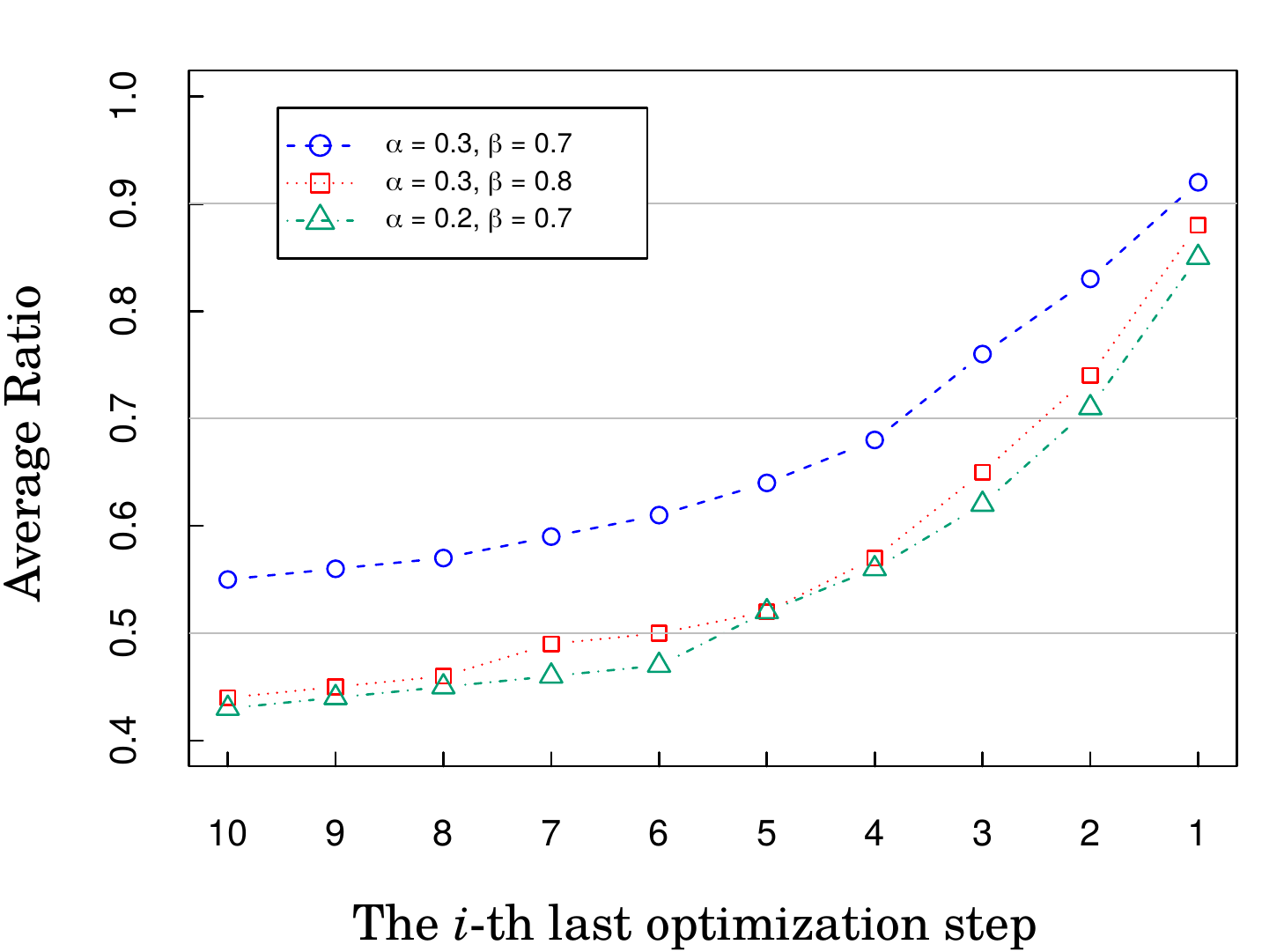}}
\caption{$L_0$ attacks are launched on 100 randomly selected images from CIFAR-10. For each of the last 10 optimization steps, we examine the average ratio $\bar{\rho}$ of the 100 intermediate distorted images. }\label{fig:adapt_steps}
\end{figure}

\vspace{3pt}
\noindent \textbf{Answer to Q1.} Let $N_\mathcal{A}$ be the number of
altered pixels and $N_\mathcal{E}$ 
the number of such altered pixels that possess \emph{extreme values} 
(i.e., values smaller than $\alpha$ or larger than $\beta$).
We consider the ratio $\rho = N_\mathcal{E}/N_\mathcal{A}$ as an indicator showing the percentage of pixels with large-amplitude perturbations, and 
want to understand how this ratio changes in the AE generation process.
As an empirical analysis, we carry out SGD step by step on 100 randomly selected images from CIFAR-10. 
For each of the last 10 steps, we calculate an average ratio $\bar{\rho}$ value of the 100 intermediate images, 
as shown in Figure~\ref{fig:adapt_steps}. We observe that the average ratio $\bar{\rho}$ goes higher and higher as $N_\mathcal{A}$ decreases. Finally, when the optimal resolution is found, around 90\% of the  altered pixels by average possess extreme values.
This helps understand why the proposed technique works well, since it is designed
to deal with such large-amplitude perturbations by recovering these pixels. 

An adversary who is aware of the details of the proposed technique
thus should try to control the amplitude of those altered pixels while satisfying the $L_0$ optimization target (i.e., minimizing $N_\mathcal{A}$). Thus, given an image, we run the SGD multiple times; once the value of $\rho$ is over 80\% (note this value finally can reach 90\% by average), we explore different optimization paths. The result shows that only 5\% of the cases succeed to control the ratio $\rho$ under 80\%. Therefore, it is difficult to control the amplitude of the altered pixels while 
satisfying the $L_0$ optimization target.

\vspace{3pt}
\noindent \textbf{Answer to Q2.} 
We follow the procedures described in \cite{he2017adversarial} to adaptively search 
potential $L_0$ AEs. Our design of the adaptive $L_0$ attack is as follows.
Since inpainting is used in both detection and defense, the adversary integrates
it into the AE generation; during the AE generation, the intermediate image
at each step of the optimization procedure
is processed using our inpainting pre-processor. Next, we check whether the resulting image is a successful attack.
If that it cannot successfully fool the neural network, we iteratively run SGD multiple times (10 in our experiments) until a resolution is found.
We randomly select 100 color images from CIFAR-10,
Our experiments show that the final number of altered pixels only takes up less than 2\% of the total number of the pixels in images from CIFAR-10, which means that they achieve the $L_0$ optimization target. 
The result finally shows that only 7\% of cases can generate successful AEs to evade 
our detection. In contrast, \cite{he2017adversarial} shows that adaptive attacks using a similar method can bypass \emph{feature squeezing}~\cite{xu2017feature} at 100\%.
Therefore, our method is much more resilient than prior work. 

\vspace{3pt}
\noindent \textbf{\emph{Summary.}} 
Based on these explorations, we conclude that $L_0$ attacks have an inherent limitation,
and it is difficult for adaptive attacks to overcome the limitation to bypass our detection.

%% file: related.tex
\section{Related Work}\label{sec:relate_work}

Generally, the protection strategies against AE attacks fall into two groups, i.e., detection and defense. In this section, we will briefly review them both.

\subsection{Detecting Adversarial Examples}

An AE detector is a binary classifier which is designed to distinguish an adversarial sample from a legitimate one. There are two strategies which are often used to design AE detectors, i.e., adversarial training and predication mismatch.

\subsubsection{Detector Training}
Some techniques use both AEs and legitimate images to train a detector. For example, Li et al.~\cite{li2017adversarial} extract PCA features after inner convolutional layers of the neural network, then use a cascade classifier to detect AEs. Metzen et al. ~\cite{metzen2017detecting} use both adversarial and benign samples to train a CNN-based auxiliary network. This light-weight sub-network works with the target model to detect AEs. They usually require a large number of samples to train the model while we only need a relatively small dataset. More importantly, our detector achieves a high detection rate but low FPR for handling $L_0$ AEs. 

\subsubsection{Prediction Mismatch}
Some techniques use the prediction mismatch strategy. For example, Bagnall et al.~\cite{bagnall2017training} train an ensemble of multiple models to use a rank voting mechanism to combine those outputs. In this way, an ensemble disagreement can be used to detect adversarial examples. \emph{Bi-model}~\cite{monteiro2018generalizable} firstly employs two pre-trained distinct models to generate features, then feeds the concatenated features to an additional binary classifier. 

Other works~\cite{tian2018detecting, meng2017magnet, xu2017feature, liang2018detecting} apply pre-processors on an input image.
For example, Meng et al.~\cite{meng2017magnet} train an auto-encoder as the image filter. Tian et al.~\cite{tian2018detecting} pre-process the images with randomly rotation and shifting since adversarial examples are usually sensitive to such transformation operations.
Liang et al.~\cite{liang2018detecting} implement an adaptive denoiser based on image entropy as the filter. 
Their methods then feed the original image and the processed one to the same neural network--if the predictions of the two images fail to match, the input is adversarial. 
In addition, Xu et al.~\cite{xu2017feature} propose \emph{feature squeezing} to detect AEs by comparing the prediction on original inputs with that on the squeezed ones. However, the proposed detector outperforms \emph{feature squeezing} for handling $L_0$ attacks. Note that the performance of their approach heavily relies on the effectiveness of the feature squeezing methods. On the contrary, our Siamese-based detector does not rely on powerful pre-processing (see Section~\ref{sec:var_pro}). 
 
Our detection technique seems close to the approaches in the second category but is actually very different. Rather than using a simple mismatch or a distance value to describe the discrepancy between an AE and its manipulated image, our technique uses a Siamese network to automatically extract the discrepancy between the two as features for detection.

\subsection{Defense}
The primary task of defensive techniques is to alleviate or eliminate the influences of AEs. 
In general, the current defensive techniques can be grouped into two major categories, that is, model enhancement and input transformation.

\subsubsection{Model Enhancement}
The first category improves the resilience of neural networks by including AEs in the process of model training, i.e., \emph{adversarial training}~\cite{goodfellow2014explaining, madry2017towards}. However, this type of defense is usually less effective against black-box attacks than white-box attacks considering the training only focuses on one certain neural network. Also, Xu et al. claim that this kind of technique suffers high cost because of iterative re-training with both adversarial and normal examples~\cite{xu2017feature}. Alternatively, \emph{defensive distillation} is proposed, and can obstruct the neural networks from fitting too tightly to the data~\cite{papernot2015distillation}. However, the prior work~\cite{carlini2017towards} demonstrates that the approach can be easily circumvented with a minimal modified attack such as a CW-$L_0$.  \emph{Shield} ~\cite{das2018shield} enhances a model by re-training it with multiple levels of compressed images based upon JPEG. However, this method is still ineffective against $L_0$ attacks.

\subsubsection{Input Transformation}
For the second category of defenses, researchers have averted their eyes from neural networks to the adversarial inputs themselves. In short, pre-processing the inputs before feeding them to networks is helpful for increasing the prediction accuracy even when facing adversarial examples. The reason why adversarial examples could successfully fool the deep learning model without being perceived is that attackers take advantage of the information redundancy of images to add adversarial noise. Consequently, well designed filters or denoisers can be considered a cure for adversarial images by removing unwanted noise. For instance, Liao et al.~\cite{liao2018defense} propose higher-level guided denoisers to remove the adversarial noise from inputs; however, their approach is computationally expensive and their work does not show its effectiveness on $L_0$ attacks. Some other methods adopt compression techniques, such as PCA~\cite{bhagoji2018enhancing} and JPEG 
~\cite{ das2017keeping, dziugaite2016study, prakash2018protecting,guo2017countering},  to filter out the information redundancy which may provide living space for adversarial perturbations in images; however, these approaches are not suitable for $L_0$ attacks. Furthermore, Bafna et al.~\cite{bafna2018thwarting} independently propose a defense against $L_0$ attacks; but their Fourier-transform-based approach is not as effective as ours (see Section~\ref{sec:eval_preproc}).

There exist some approaches that do not fall in either category. For example, 
MVP-Ears~\cite{zeng2019multiversion} borrows the idea of multi-version programming from software engineering 
and applies it to audio AE detection. It deploys
multiple diverse automatic speech recognition systems in parallel, and detects audio AEs by comparing their recognition
results. However, the idea will probably fail in handling image AEs, which are known to have good transferability~\cite{goodfellow2014explaining}.

%% file: conclude.tex
\section{Discussion}\label{sec:limit}

First, the proposed technique is not a panacea for detecting or defending against all possible attacks. 
As future work, we are to explore whether the Siamese network-based detector can be generalized to detect other types
of attack, such as $L_{\infty}$. 

Second, our adaptive attack (following the procedures in~\cite{he2017adversarial}) is based on the exploration of 
different optimization paths. 
There exist some other alternative white-box attacks, such as the method proposed in~\cite{carlini2017adversarial} which attempts to create new AEs by modifying the loss functions to bypass detectors.  
Whether other different adaptive attacks, such as~\cite{carlini2017adversarial}, can bypass our detector is interesting, and we plan to investigate it in the future.

Third, how to prevent the over-fitting problem is still an open question in the field of machine learning. Although we took over-fitting into consideration when designing the experiments (e.g., the testing dataset and training dataset are completely disjoint), the over-fitting problem is still possible. Specifically, our evaluation only examined 
base classifiers ResNet and CNN. Future work will consider other classifiers
with different architectures.


Fourth, we are also interested in whether the proposed technique is scalable to handle a larger data set such as ImageNet. We leave this evaluation as the future work.

Lastly, this work shows that only controlling the number of altered pixels 
without limiting the resulting amplitude weakens the power of the generated AEs. Thus, how to make a good trade-off between the number of altered pixels and their amplitude 
becomes critical when designing new AE generation algorithms.

\section{Conclusions}\label{sec:conclude}

In the setting of classic $L_0$ AE attacks, a bounded number of pixels can be corrupted without limiting the amplitude. These large-amplitude pertubations in $L_0$ AEs are considered as a challenge by many previous works, since the effect of such corruptions is difficult to eliminate. Considering the threats caused by $L_0$ AEs, a highly accurate detection technique and an effective
defense that can rectify the classification under $L_0$ perturbations
are urgently needed. By identifying and exploiting the inherent characteristics of $L_0$ AEs, 
we develop \textsc{AEPecker} that thwarts this type of attacks. 
Its novel Siamese network based design shows very high accuracies
in detecting $L_0$ AEs, and its inpainting-based preprocessing technique
can effectively rectify those AEs and thus correct the classification
results. Plus, it is resilient to adaptive attacks that bypass
prior approaches.
